\documentclass[%
 reprint,
 amsmath,amssymb,
 aps,
]{revtex4-2}

\usepackage{graphicx}
\usepackage{dcolumn}
\usepackage{bm}
\usepackage{xcolor}
\usepackage{appendix}

\begin{document}

\preprint{APS/123-QED}

\title{Observations of rogue seas in the Southern Ocean}

\author{A. Toffoli$^{1}$}
\author{A. Alberello$^{2}$}
\author{H. Clarke$^1$}
\author{F. Nelli$^{3}$}
\author{A. Benetazzo$^4$}
\author{F. Bergamasco$^5$}
\author{B. Ntamba Ntamba$^6$}
\author{M. Vichi$^{7,8}$}
\author{M. Onorato$^{9,10}$}

\affiliation{$^1$Department of Infrastructure Engineering, The University of Melbourne, Parkville, VIC 3010, Australia;}
\affiliation{$^2$School of Mathematics, University of East Anglia, Norwich, United Kingdom;}
\affiliation{$^3$Department of Mechnaical Engineering, Swinburne University of Technology, Melbourne. Australia}
\affiliation{$^4$Istituto di Scienze Marine, Consiglio Nazionale delle Ricerche, 30122 Venice, Italy}
\affiliation{$^5$Universit{\`{a}} C{\'{a}} Foscari, 30123 Venice, Italy}
\affiliation{$^6$Cape Peninsula University of Technology, 7535 Cape Town, South Africa}
\affiliation{$^7$Department of Oceanography, University of Cape Town, Cape Town, South Africa}
\affiliation{$^8$Marine and Antarctic Research centre for Innovation and Sustainability, University of Cape Town, Cape Town, South Africa}
\affiliation{$^9$Dipartimento di Fisica, Universit{\`a} degli Studi di Torino, Via Pietro Giuria 1, 10125 Torino, Italy}
\affiliation{$^1{^0}$INFN, Sezione di Torino, Via Pietro Giuria 1, 10125 Torino, Italy}

\date{\today}

\begin{abstract}
{We report direct observations of surface waves from a stereo camera system along with concurrent measurements of wind speed during an expedition across the Southern Ocean in the austral winter aboard South African icebreaker S.A.~Agulhas~II.} Records include water surface elevation across a range of wave conditions, spanning from early stages of wave growth to full development. We give experimental evidence of rogue seas, i.e., sea states characterizided by heavy tails of the probability density function well beyond the expectation based on bound mode theory. These conditions emerge during wave growth, where strong wind forcing and high nonlinearity drive wave dynamics. Quasi-resonance wave-wave interactions, which are known to sustain the generation of large amplitude rogue waves, capture this behaviour. Wave statistics return to normality as the wind forcing ceases and waves switch to a full developed condition.

\end{abstract}

\maketitle 

{Ocean surface waves result from momentum transfer and energy exchange driven by wind-induced surface pressure} \cite{janssen2004interaction}. Under the direct action of atmospheric forcing, waves are generated and grow in height as a function of fetch, i.e.,\ the distance over which wind blows unobstructed. {Because of the nonlinearity of the free surface, the energy and wave action  
are then redistributed among the Fourier modes \cite{zakharov1968stability}. The peak of the wave spectrum downshifts, so that the dominant waves become longer and faster}. {Additionally, there is transfer of energy across directions, imparting} a distinctive two-dimensional (directional) nature to the wave spectrum \cite{janssen2004interaction, toffoli2017wind}. Wave growth ceases when waves become faster than the wind, a condition which normally occurs when the wave age, i.e.,\ the ratio of wave phase velocity ($C_P$) to wind speed ($U$), exceeds 1.25 \cite{janssen2004interaction}. 

If the wave steepness---a parameter proportional to the ratio of wave height to wavelength---is sufficiently small, the resulting ocean surface can be considered as a superposition of many wave components, and its statistical features can be approximated by a Gaussian distribution \cite{janssen2004interaction,toffoli2017wind,onorato09}. 
In nature, however, waves are steep, and the small amplitude assumption does not always hold. Consequently, nonlinear wave-wave interactions can develop, {enhancing the likelihood of large amplitude waves and causing deviations from Gaussian statistics \cite{janssen2003nonlinear,onorato09,waseda2009evolution,mori2011estimation}.} 
{A proxy for these is the kurtosis of the surface elevation \cite{fedele2015kurtosis,onorato09,mori2011estimation}, which is the fourth-order moment of its probability density function.}

{There are two types of nonlinear interactions contributing to non-Gaussian behaviours. The first, attributed to bound modes, introduces waveform asymmetry, and skews the distribution of surface elevation \cite{tayfun1980narrow,forristall2000wave,toffoli2007second,janssen2009some}. This results in a discernible, albeit weak, increase in kurtosis relative to a Gaussian random process \cite{tayfun1980narrow}. A second, and more intense, type arises from quasi-resonant interactions between free modes, which can be considered as a generalization of the modulational (or Benjamin-Feir) instability for a random wave field \cite{onorato2013rogue,dudley2019rogue}. 
This has the potential to causes the emergence of heavy tailed statistics \cite{janssen2003nonlinear,mori2006kurtosis,onorato09,waseda2009evolution,fedele2015kurtosis,walczak2015optical} through the sudden appearance of rogue waves, which are notably large events amidst smaller waves.}
{Quasi-resonant interactions are effective when waves are steep, the wave spectrum is sufficiently narrow-banded and wave propagation is predominantly unidirectional \cite{onorato09}. However, the typical } spectrum of ocean wind waves is broad-banded in frequencies and spread over multiple directions \cite{teutsch2023rogue,derkani_essd_2021}. {Therefore,} the capacity to develop modulational instability seems to be impaired in natural conditions \cite{mori2011estimation,fedele2016real}. Under these circumstances, theory, experiments, and numerical simulations showed that wave statistics exhibit weakly non-Gaussian properties, deriving primarily from bound modes \cite{onorato09,waseda2009evolution,socquet2005probability}. This is further supported by field observations \cite{forristall2000wave,toffoli2007second}, even though rogue waves are present in the records \cite{mori2002analysis,donelan2017making,gemmrich2022generation,fedele2016real,bitner2014north,benetazzo2015observation}.

{Nevertheless, the degree to which wind forcing impacts wave statistics has received limited attention. Yet, there is numerical and experimental evidence that wind can drive the growth and sustain the lifetime of large amplitude waves through higher wind-induced pressure drag \cite{kharif2008influence,leemontywindwaves2020}, enhance the modulational instabily \cite{onorato2012approximate,brunetti2014modulational}, and increase deviations from Gaussian statistics \cite{toffoli2017wind,leemontywindwaves2020}. Therefore, it remains unclear whether strong non-Gaussian behaviors,} well beyond those associated with bound modes, can occur in the ocean \cite{fedele2016real}{, where the atmospheric forcing is an inherent factor}. {Consequently, the statistical} properties of ocean waves in general, and the occurrence of extreme waves in particular, remain the subject of controversy, making inclusion of extreme and rogue waves in design considerations a challenging endeavor \cite{bitner2014occurrence,BITNERGREGERSEN201426}. 

\begin{figure}
\includegraphics[width=0.48\textwidth]{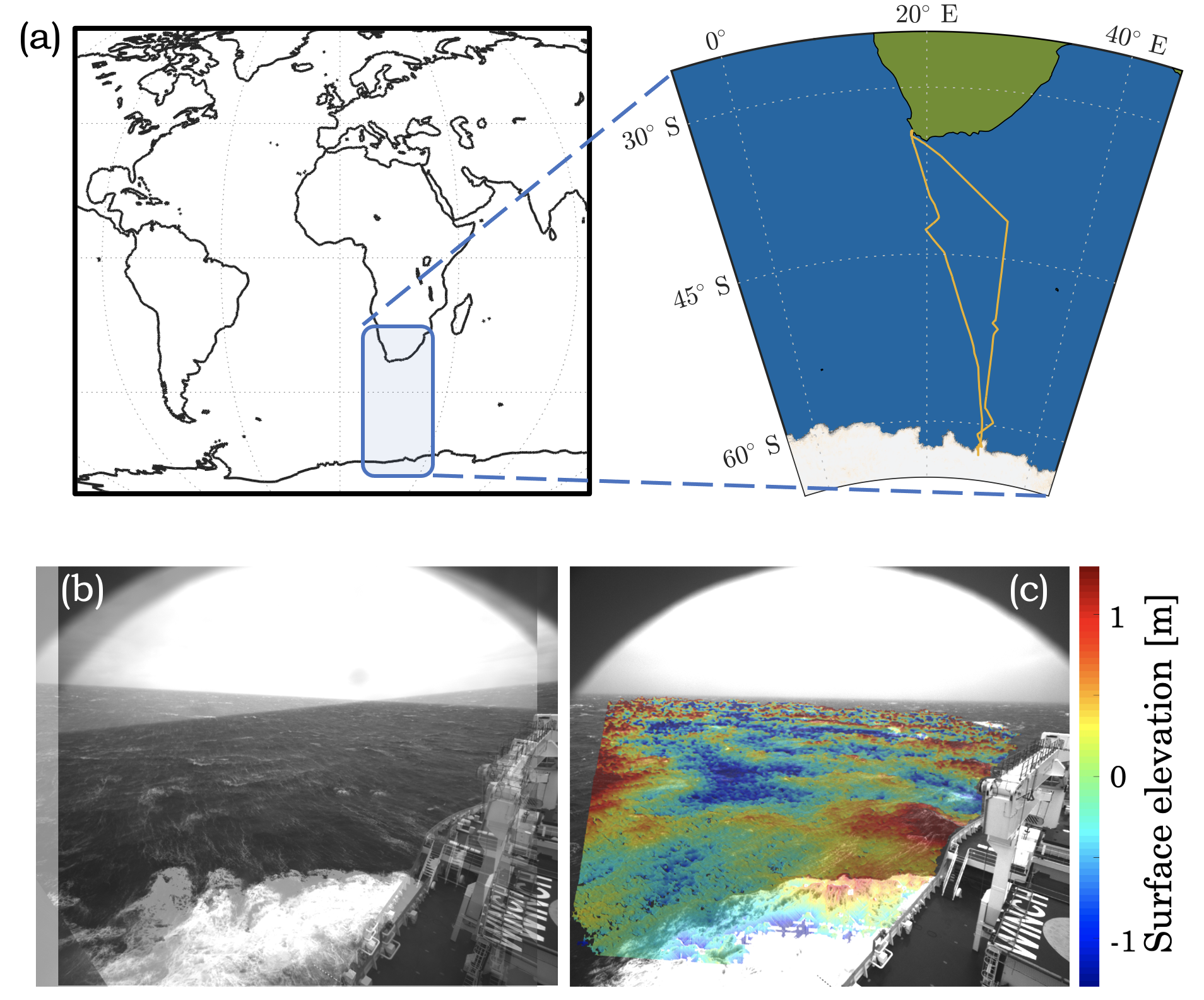}
\caption{\label{fig:3d}Ship track (yellow lines; a) and sample images (b,c): example of overlapping stereo images (b) and reconstructed three dimensional surface (c)}
\end{figure}

{In this Letter, we present field (in-situ) observations of concurrent ocean surface and lower atmosphere properties, encompassing diverse sea states and wave growth phases, to reveal the extent of which the wind forcing, coupled with nonlinear interactions, contributes to the emergence of heavy-tailed statistics.}
Measurements {of the ocean surface} were acquired underway by a stereo camera system aboard the South African icebreaker S.A.~Agulhas~II during a crossing of the Southern Ocean in the austral winter (28~June---13~July, 2017; Fig.\ref{fig:3d}a) \cite{alberello2019observation,vichi2019effects}. This remote region comprises an uninterrupted band of water around Antarctica \cite{young2020wave}. It is dominated by strong westerly winds, the notorious roaring forties, furious fifties and screaming sixties, which give rise to some of the fiercest waves on the planet over almost infinite fetches all the year round \cite{alberello2022three,young2020wave,derkani_essd_2021,barbariol2019large}. {As a reference, the significant wave height---an average measure of the highest third of waves in a given sea state and a proxy for wave height statistics in the ocean---has a 50-th percentile of $\approx$ 5~m and a 90-th percentile of $\approx$ 7~m in the austral winter \cite{young2020wave}.} 

The acquisition system consisted of two synchronised GigE monochrome industrial CMOS cameras with a 2/3 inch sensor, placed side-by-side at a distance of 4\,m. The stereo rig was installed on the monkey bridge of the icebreaker approximately 25~m from the waterline and tilted 20$^{\circ}$ below the horizon. The cameras were equipped with 5\,mm lenses to provide a field of view of the ocean surface $\approx{}90^{\circ}$ around the port side of the ship. Additionally, an inertial measurement unit was mounted close to the two cameras to capture their movement with respect to the sea surface. Images were recorded at a sampling rate of 2\,Hz with a resolution of 2448$\times$2048 pixels during daylight, and were partitioned in sequences of 30 minutes. 
Complementing atmospheric data {from the ship's meteorological station} as well as navigation records {from the ship's voyage data recorder} were acquired throughout the journey. {Two wave buoys were also deployed during the expeditions to provided supplementary records of the surface elevation \cite{alberello2022three,vichi2019effects}.}

\begin{figure}
\includegraphics[width=0.48\textwidth]{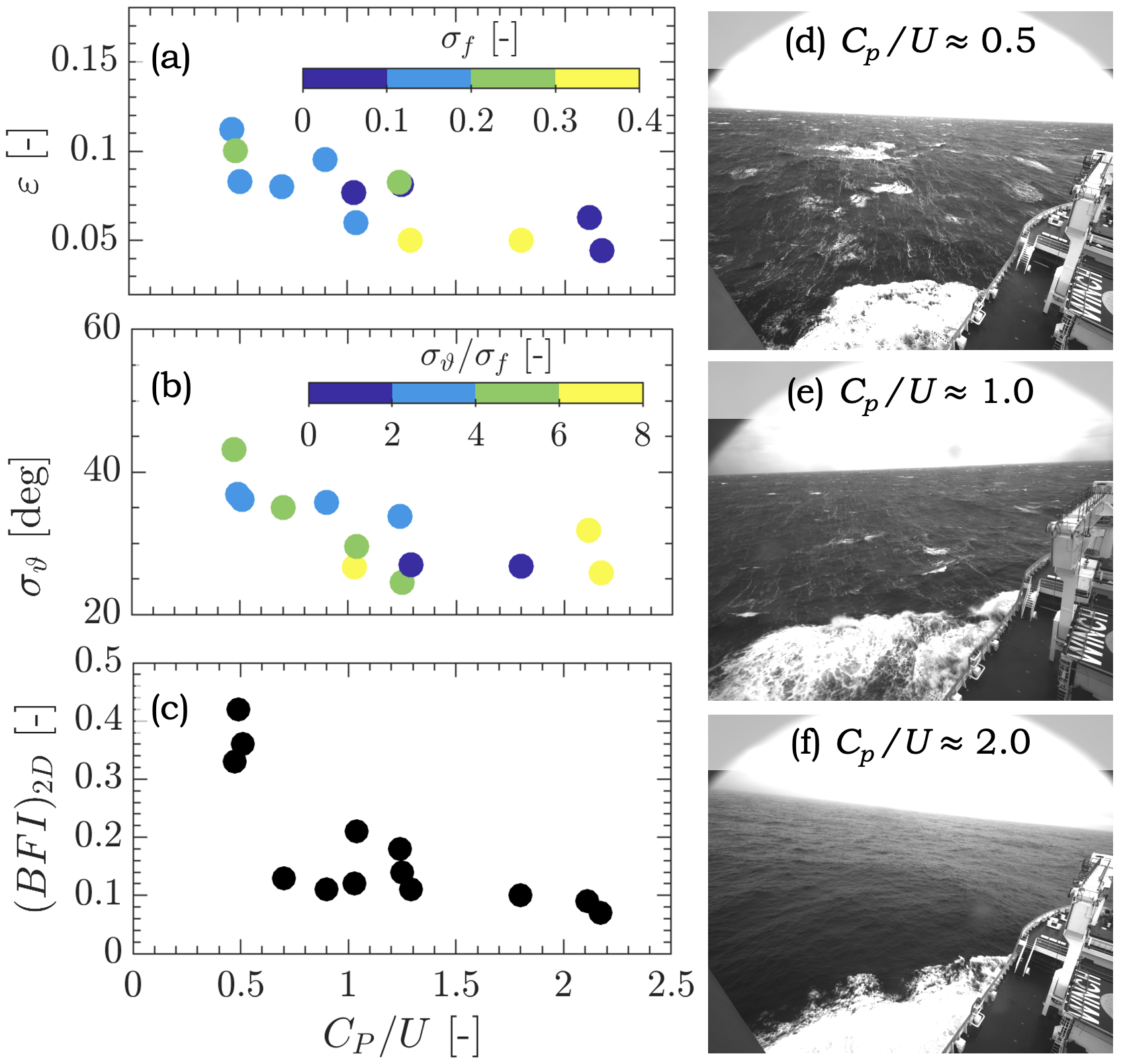}
\caption{\label{fig:par}Evolution of spectral characteristics across different stages of wave growth as measured by the wave age: (a) wave steepness (symbols) and frequency bandwidth (color map); (b) directional spreading (symbols) and directional spreading relative frequency bandwidth (color map); and two-dimensional Benjamin-Feir Index. Examples of sea state images at different wave age are reported: (d) growing (young) sea state ($C_P / U \approx 0.5$); (e) near fully developed sea state ($C_P / U \approx 1$); and a fully developed sea state ($C_P / U \approx 2$).}
\end{figure}

Pairs of synchronized images were processed to reconstruct the three-dimensional ocean surface displacement. {Accuracy relies on sensor's capacity in capturing, processing, and displaying image-forming signals. In the open ocean, this is influenced by environmental conditions that alter light exposure and shadowing effects in the lee of large waves that conceal parts of the ocean surface. Hereafter, we discuss a selection of sequences recorded in the open ocean with optimal light conditions and low uncertainties, (see details on method and accuracy of the measurements in Appendix A).} 

An example of stereo pair and reconstructed surface is shown in Fig.~\ref{fig:3d}b,c. From each sequence, the directional wave spectrum was derived (see details in \cite{alberello2022three}), and spectral parameters were computed. These are reported in Fig.~\ref{fig:par} as a function of the wave age relative to the spectral peak. Parameters include: the steepness, which is expressed as $\varepsilon = k_P H_s / 2$, where $k_P$ is the wavenumber at the spectral peak, and $H_s$ is the significant wave height, computed as $4\sqrt{m_0}$, with $m_0$ being the zeroth order moment of the wave spectrum; the frequency bandwidth $\sigma_f$, which refers to the half width at half maximum of the dominant spectral peak relative to its peak frequency, noting that the restriction to the most energetic peak avoids ambiguities arising from the coexistence of multiple wave systems;
the directional spreading $\sigma_{\vartheta}$, which is the circular standard deviation of the directional spectrum; and the two dimensional Benjamin-Feir Index $(BFI)_{2D}$ \cite{mori2011estimation}, which is calculated as 
\begin{equation}
(BFI)_{2D} = \frac{\sqrt{2} \, \varepsilon}{\sigma_f}\,\left[\frac{1}{1+7\,R}\right],
\label{bfi2d}
\end{equation}
with $R = 0.5 (\sigma_{\vartheta}/\sigma_f)^2$.
Among these parameters, the steepness is a general indicator for wave nonlinearity. Its effect on the kurtosis can be expressed as \cite{mori2006kurtosis}
\begin{equation}
(\kappa)_{Bound} =  3 + 24 k_P^2 m_0,
\label{kurt_bound}
\end{equation} 
and it represents the contribution of bound waves \cite{tayfun1980narrow}.
The $(BFI)_{2D}$ is a proxy for quasi-resonant interactions between free modes \cite{onorato2013rogue,mori2011estimation}. Their effect on the kurtosis can be calculated as \cite{mori2011estimation}
\begin{equation}
(\kappa)_{Free} =  3 + \frac{\pi}{\sqrt{3}}\,(BFI)_{2D}^2,
\label{kurt_free}
\end{equation}
and significant contributions can be expected when $(BFI)_{2D} \gtrapprox 1$. The overall kurtosis ($\kappa_T$) is the sum of bound and free wave contributions.  

For young sea states ($C_P/U < 1$), the significant wave height varied within 3.5--4.5~m and wave period ranged 9--10~s{, noting that uncertainties related to the stereo image processing can induce errors of $\approx5\%$ for the former and $\approx2.5\%$ for the latter \cite{alberello2022three,benetazzo2016stereo} relative to collocated buoy measurements.} The waves were steep, with $\varepsilon > 0.08$ (Fig.~\ref{fig:par}a), which is typical during intense storms \cite{toffoli2005towards} and reported during observations of extreme waves in the laboratory and in the field \cite{onorato09,toffoli2017wind,bitner2014north,gemmrich2022generation}. However, the spectral shape was relatively broad-banded, both in the frequency and directional domain (Fig.~\ref{fig:par}a,b). As the latter dominated on the former (i.e.,\ $\sigma_{\vartheta} / \sigma_f > 1$; Fig.~\ref{fig:par}b), it follows from equation (\ref{bfi2d}) that $(BFI)_{2D} \approx 0.4$ (Fig.~\ref{fig:par}c). Nevertheless, despite weak quasi-resonant interactions, the sea state retained an active nonlinearity. Consistently with laboratory experiments and numerical simulations \cite{waseda1999experimental,iafrati2013modulational}, this was demonstrated by the occurrence of whitecaps (formation of frothy, aerated crests indicating occurrence of wave breaking dissipation; Fig.~\ref{fig:par}d) or, in some instances, multiple whitecaps recurring from the same wave train (see supplementary video [URL will be inserted by publisher]). As the sea state evolved into more mature conditions ($C_P/U \approx 1$), and eventually reached a fully developed stage ($C_P/U > 1.25$), the wave steepness gradually reduced (significant wave height ranged 2--3~m and wave period ranged 10--14~s). 
It should be noted that two cases of broad spectra are reported in Fig.~\ref{fig:par}a for $C_P/U > 1$. These sea conditions are attributed to the coexistence of multiple {wave systems} of similar peak frequency arising from rapidly veering wind. These peaks merge into a single broader spectral peak, and thus go undetected by an automated analysis. Similarly to the spectral bandwidth, also the directional spreading contracted to a smaller range, in agreement with laboratory experiments \cite{toffoli2017wind} (Fig.~\ref{fig:par}a,b). The $(BFI)_{2D}$ showed a distinctive decrease from $\approx$ 0.4 to $\approx$ 0.1 (Fig.~\ref{fig:par}c) due to the drop in steepness and concurrent increase of $\sigma_{\vartheta} / \sigma_f$. Therefore, as wind forcing ceased, the nonlinear nature of the sea state faded. This was further substantiated by limited, if any, occurrence of whitecaps (Fig.~\ref{fig:par}e,f).

\begin{figure}
\includegraphics[width=0.45\textwidth]{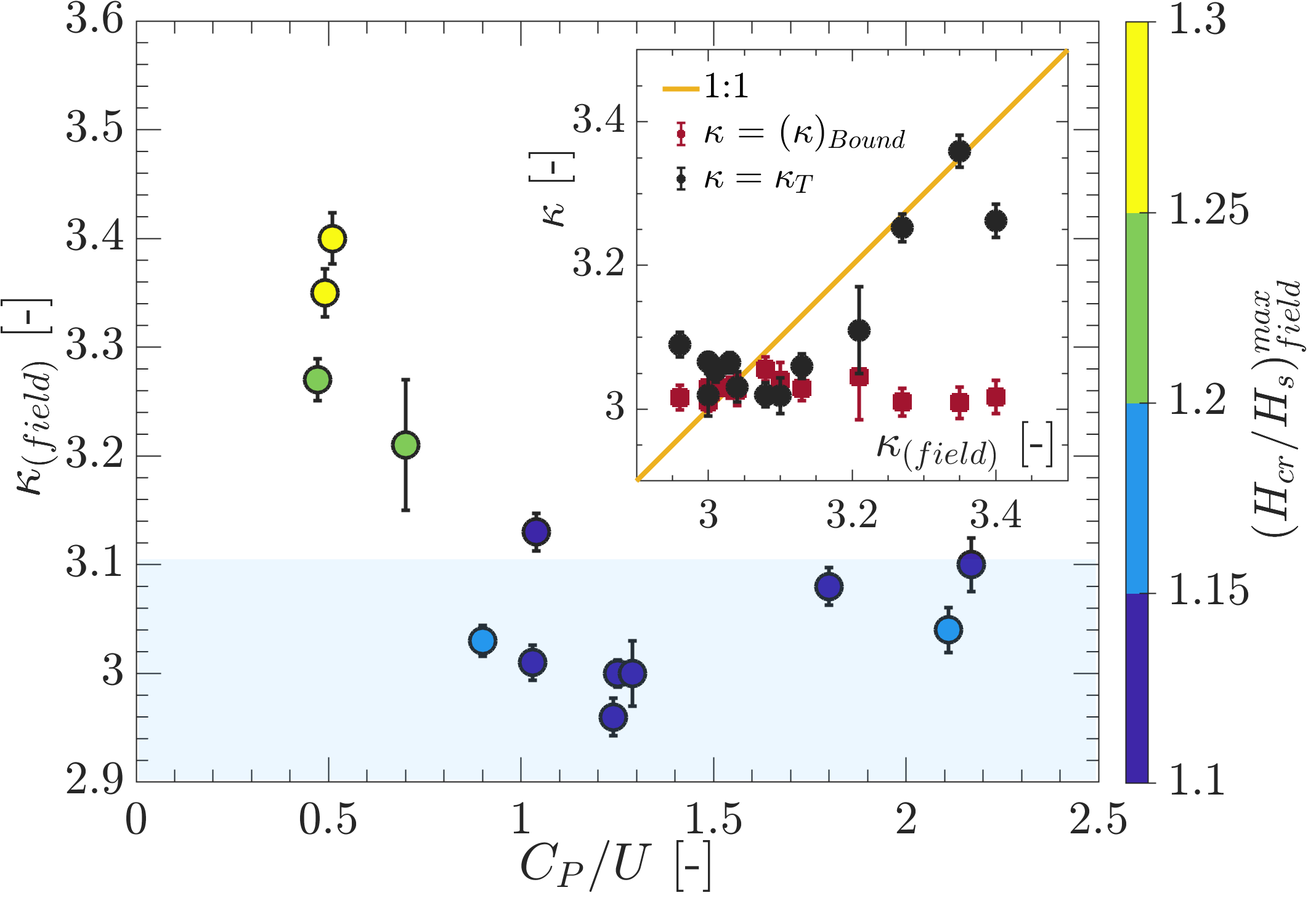}
\caption{\label{fig:kurt}Field observations of the kurtosis of the surface elevation as a function of wave age. Estimates of the maximum wave crest based on a fitted three-parameter Weibull distribution are reported as a color map. The (light blue) shaded area is shown to identify values of kurtosis normally expected for Gaussian and weakly non-Gaussian sea states. A comparison against theoretical predictions based on bound and free wave contributions is shown in the inset. The error bars represent the 95\% confidence interval.}
\end{figure}

\begin{figure*}
\includegraphics[width=\textwidth]{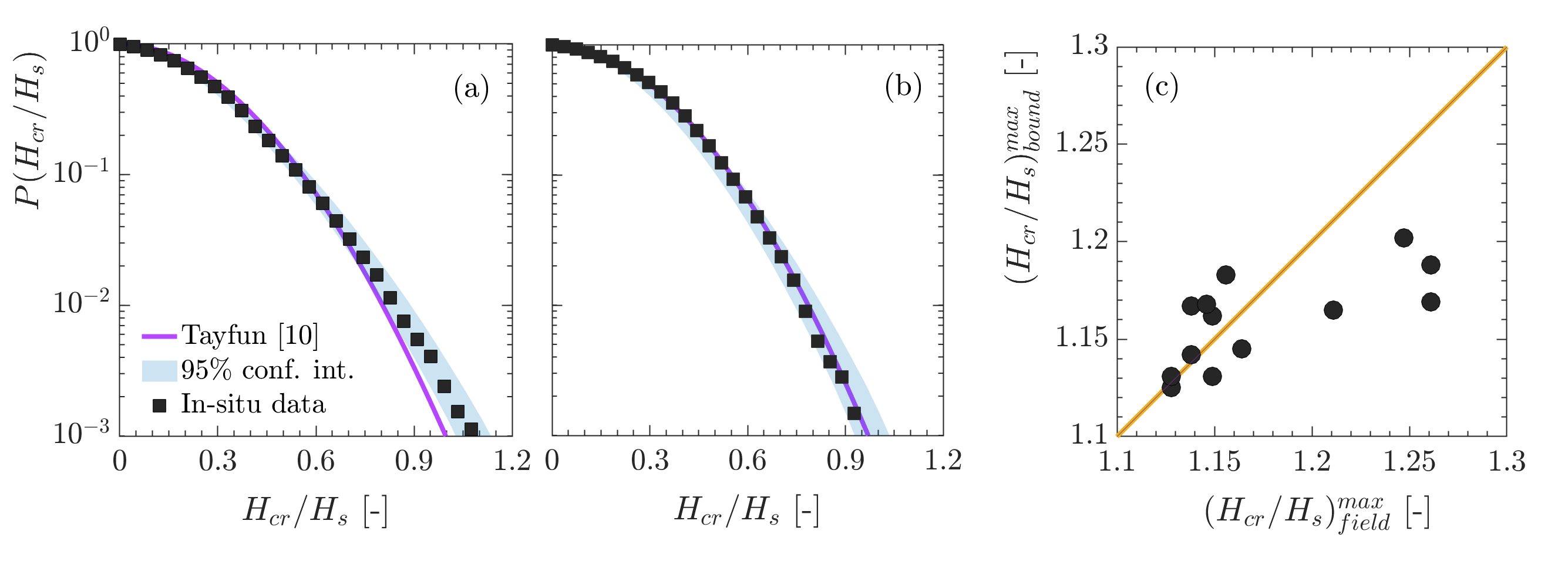}
\caption{\label{fig:pdf_2nd}Wave crest distribution in young ($C_P / U \approx 0.5$; a) and fully developed ($C_P / U \approx 2$; b) sea states from in-situ data and theory \cite{tayfun1980narrow}. Comparison of maximum wave crest amplitudes extrapolated from in-situ data and theoretical \cite{tayfun1980narrow} distributions (c).}
\end{figure*}

Wave statistics is discussed in the form of the kurtosis of the surface elevation, and it is presented in Fig.~\ref{fig:kurt} as a function of wave age. Note that the kurtosis computed from observations include both bound and free wave contributions. For completeness, the maximum wave crest relative to significant wave height is also reported. In contrast to a direct estimation of the maximum crest height in a sequence (i.e.,\ maximum wave elevation), we extrapolated it from a Weibull distribution fitted to the experimental data (see Appendix B), to allow further comparison with theoretical counterparts. 

The kurtosis for a Gaussian or weakly non-Gaussian sea state is normally expected within 2.9--3.1 (e.g.\ \cite{onorato09}). For young sea states ($C_P / U \approx 0.5$), the kurtosis departed from this benchmark and reached values as high as 3.4. {Small errors in wave height due to factors affecting image acquisition and processing have minimal impact. Accounting for these errors in equations (\ref{kurt_bound},\ref{kurt_free}) or in surrogate time series results in variations within 1-2\% in kurtosis. The latter is also susceptible to instability due to sample size \cite{toffoli2010evolution}. The error bars in Fig.~\ref{fig:kurt} quantify this uncertainty, representing twice the standard deviation of kurtosis from 1,000 random sample surrogates. This attributes a 95\% confidence interval width of $10^{-2}$.}
In the context of water waves, the most extreme events are normally restrained by breaking (see Fig.~\ref{fig:par}d), which {prevents
excessive increase in kurtosis}. In this regard, similar magnitudes of kurtosis were reported in laboratory observations of highly nonlinear mechanically and wind-generated water waves \cite{onorato09,waseda2009evolution,toffoli2017wind}. Comparison with theoretical estimates from equations (\ref{kurt_bound},\ref{kurt_free}) indicates that bound wave nonlinearity cannot capture strongly non-Gaussian statistics, while quasi-resonance interactions can detected them to a certain extent (see inset in Fig. \ref{fig:kurt}). As the sea state developed into more mature conditions, the kurtosis decreased exponentially, approaching Gaussian statistics already for $C_P / U \approx 1$. Theoretical estimates (\ref{kurt_bound},\ref{kurt_free}) are consistent with observations under these circumstances, substantiating the weak nonlinear properties of well developed sea states. 

Consistently with the high value of the kurtosis in young seas, the estimated maximum crest height, relative to the significant wave height, was large and in excess of 1.2, which is the threshold identifying rogue waves \cite{christou2014field}. Concurrently, the recovery of Gaussian statistics for mature sea states coincided with a reduction of maximum crests, which did not exceed the rogue wave limit. 

Bound waves are expected to contribute to the wave crest statistics \cite{onorato09}. A more detailed comparison against bound wave driven statistics, which, we recall, are the benchmark statistical properties for oceanic sea states, is reported in Fig. \ref{fig:pdf_2nd}. The Tayfun distribution for wave crests \cite{tayfun1980narrow} is assumed as reference for bound wave statistics, and it is applied to extrapolate theoretical prediction of maximum crest heights {through equations (\ref{extreme_fun},\ref{extr_gumbel}) in Appendix B} \cite{toffoli2008surface}. For intense wind forcing ($C_P / U < 1$), the in-situ wave crest distribution shows a moderate, yet statistically significant (i.e., in excess of the 95\% confidence intervals), deviation from the theoretical counterpart (Fig. \ref{fig:pdf_2nd}a). This differs from currently available field observations, which report good agreement with bound wave statistics \cite{forristall2000wave,toffoli2007second,gemmrich2022generation}. It follows that bound waves tend to under estimate the amplitude of the most extreme crests during the growing phase of wind-generated fields (see data point referring to relative crest height in the field greater than 1.2 in Fig. \ref{fig:pdf_2nd}c). As the sea state evolves into a fully developed form, the wave crest statistics match the Tayfun distribution (Fig. \ref{fig:pdf_2nd}b) and extreme crest amplitudes reported in the field become consistent with theoretical predictions (see data point for relative wave crest height in the field lower than 1.2 in Fig. \ref{fig:pdf_2nd}c). 

\begin{figure}
\includegraphics[width=0.48\textwidth]{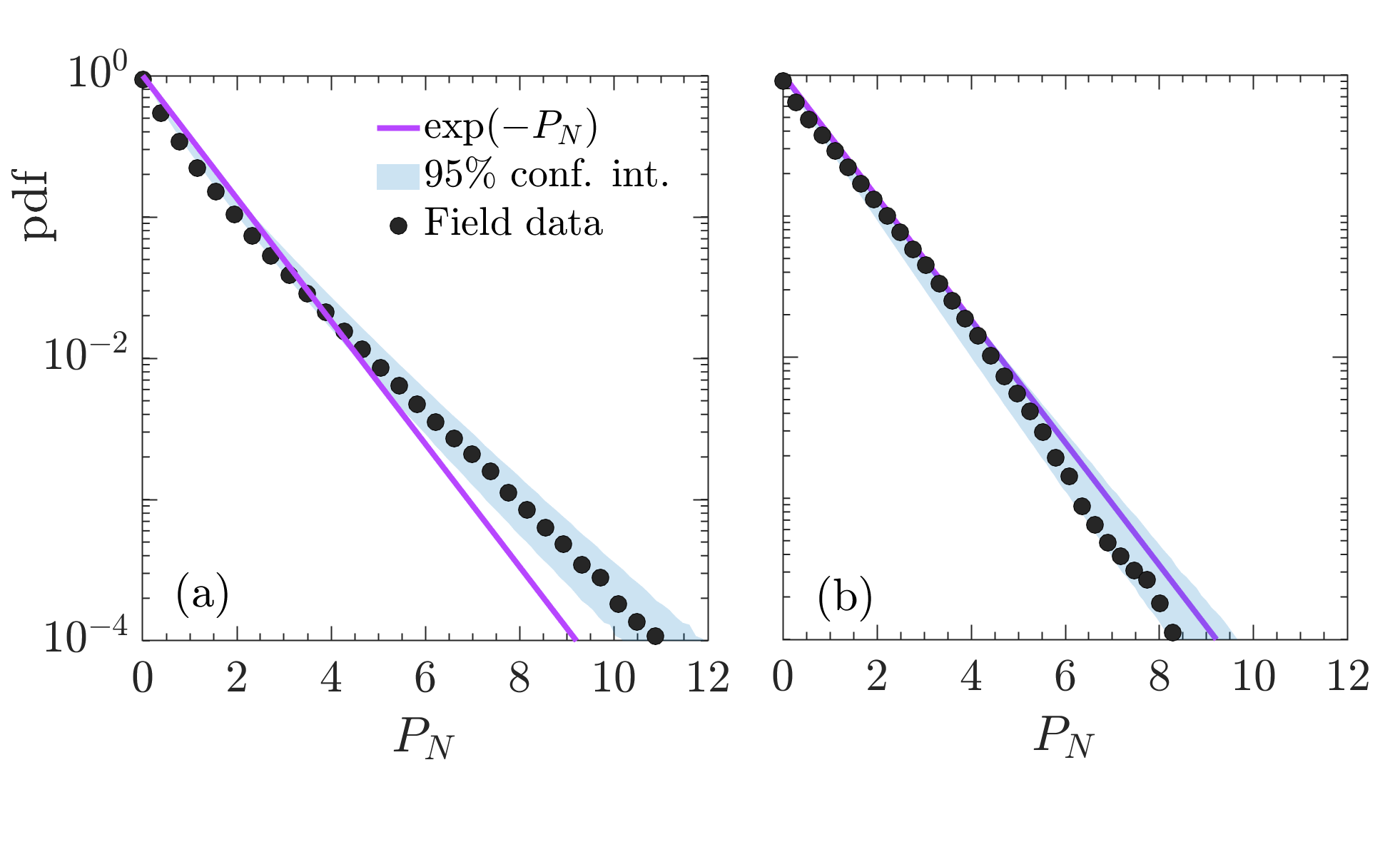}
\caption{\label{fig:pdf}Probability density function of the normalised wave intensity ($P_N$) for a young (a) and mature (b) seas; the 99\% confidence interval is indicated as shaded (light blue) area. The young sea represents data from a sequence of image acquired for wave age $C_P/U \approx 0.5$; the mature sea represents data from a sequence of image acquired for wave age $C_P/U > 1$. The $\exp(-P_N)$, which represents a Gaussian process, is reported for benchmark.}
\end{figure}

For completeness, Fig.~\ref{fig:pdf} shows the probability density functions of the normalised wave intensity $P_N$, which is the square modulus of the wave envelope normalised by its average. This is preferred to the wave displacements (surface elevation), as it better represents the upper tail of the distribution \cite{janssen2014random}. Furthermore, the wave envelope encompasses the whole reconstructed ocean surface, unlike the point approach used for bound wave statistics. For young seas (Fig.~\ref{fig:pdf}a), a significant departure from the distribution of $P_N$ expected for Gaussian statistics, i.e.\ $\text{exp}(-P_N)$ \cite{toffoli2017wind}, is evident, confirming the robustness of the heavy tail. For mature sea conditions (Fig.~\ref{fig:pdf}b), the distribution aligns to $\text{exp}(-P_N)$ more closely, substantiating the Gaussian nature of fully developed wave fields. Nevertheless, it is worth mentioning that $\text{exp}(-P_N)$ tends to slightly over estimate the distribution of in-situ data, a feature which is attributed to the broad-banded nature of the wave fields \ \cite{mori2002analysis}.  

In summary, we presented a set of in-situ observations of ocean surface displacements {derived from stereo imaging and complemented by concurrent data of atmospheric forcing.} Records encompass a wide range of sea states from the Southern Ocean, spanning from the early stages of wave development (young seas), where wave dynamics are driven by wind forcing, to fully developed stages (mature seas), where wave dynamics are no longer under the direct effect of wind. {As oceanic waves are steep, their statistical features are driven by bound waves and the interaction of free waves \cite{janssen2009some,fedele2015kurtosis}. The former are generally more persistent than the latter, resulting in weakly non-Gaussian behaviours. However, the action of strong winds during early stages of wave growth generates peaked spectra and revives the interaction of free waves, prompting heavy-tailed statistics with strongly non-Gaussian nature through a more frequent occurrence of rogue waves, which has only been previously observed in controlled laboratory experiments where waves were mechanically generated.} {We note that the accuracy of our observations is contingent on image acquisition and processing, which may vary considerably across sequences. When compared to conventional instruments like wave buoys, discrepancies in estimating significant wave height were $\approx$5\%. While this discrepancy is noteworthy, its impact on the extent of non-Gaussian properties remains minimal. Effects related to the sample size are also minor. The observations presented herein are a robust evidence that ocean waves can develop into a strongly non-Gaussian process, as theory would predict, when actively forced by wind. Our results challenge existing perspectives on rogue wave statistics and underscore the need for a more thorough consideration of wind forcing---an inherent feature in natural conditions---when predicting rogue seas.}


\section*{Acknowledgments}
The expeditions were funded by the South African National Antarctic Programme through the National Research Foundation. AA and AT were funded by the ACE Foundation and Ferring Pharmaceuticals and the Australian Antarctic Science Program (project 4434). AT is supported by the Australia Research Council (DP200102828, LP210200927). AA acknowledges the London Mathematical Society (Scheme 5, 52206) and EPSRC grant EP/Y02012X/1. MV was supported by the NRF SANAP contract UID118745. M.O. was funded by Progetti di Ricerca di Interesse Nazionale (PRIN) (2020X4T57A and 2022WKRYNL) and by the Simons Foundation (Award 652354). We are indebted to Captain Knowledge Bengu and the crew of the SA~Agulhas~II for their invaluable contribution to data collection. We acknowledge Mr G. Passerotti for helping with the supplementary material and Dr L. Fascette for technical support.

%

\section*{APPENDIX A: IMAGE PROCESSING}
The open source Wave Acquisition Stereo System (WASS \cite{bergamasco2017compgeosc}) was used to analyse simultaneous pairs of images to find photometrically distinctive corresponding points that can be triangulated to recover their original three dimensional position in space \cite{benetazzo2006measurements}, i.e.,\ the actual surface elevation. {For each sequence of images, WASS automatically estimates the geometrical configuration of the two cameras (extrinsic parameters), enabling it to detect and compensate for errors resulting from the reciprocal orientation between sensors caused by ship vibrations. The motion of the vessel under the effect of waves is more significant than the latter. Hence, three dimensional point clouds from different pairs of images lie in different reference frames after stereo rconstruction. Measurements of ship motion from the IMU are therefore used to align} and geo-localise data points to a common horizontal plane representing the mean sea level. The operation resolved an area of approximately 200~m $\times$ 200~m for each stereo image, producing more than two million data points for a sequence.

{The accuracy of the reconstructed surface elevation depends on various factors, including image resolution, settings of the stereo rig, light conditions, photographic grain effects, environmental conditions (such as air turbulence or water droplets), and lack of visibility in the lee of large waves (shadowing effect) \cite{benetazzo2016ceng,benetazzo2016stereo}. A direct comparison of wave spectra from the reconstructed surface elevation against collocated data from wave buoys (see details in \cite{alberello2022three}) shows good agreement for modes around the spectral peak. Discrepancies arise in the lower ($f < 0.05$,Hz) and upper ($f > 0.10$,Hz) tails, where overestimation and underestimation of energy relative to the buoys are reported, respectively. As the spectral tails carry a small amount of energy, the contributions to uncertainties of wave properties are minor. Overall, the significant wave height from stereo images underestimates buoy data by $\approx5$\% (cf. also \cite{benetazzo2016stereo}); the mean period is less prone to uncertainties and it differs from buoys by $\approx2.5$\%.}


\section*{APPENDIX B: MAXIMUM CREST HEIGHT} 
The maximum crests height relative to the significant wave height was extrapolated from a surrogate statistical framework (see e.g.\ \cite{toffoli2008surface}). This was achieved by first extracting individual wave crests from time series of the surface elevation at a central location in the reconstructed image domain. A three-parameter Weibull function was then fitted to the population of observed wave crests and applied to derive a distribution of extremes as
\begin{equation}
\label{extreme_fun}
F_E(H_{cr})=\left[ 1- \mathrm{exp} \left( - \; \frac{H_{cr} - \gamma_w}{\alpha_w} \right)^{\beta_w} \right]^n,
\end{equation}
where the term within the square brackets is the three-parameter Weibull distribution, $H_{cr}$ is a generic wave crest, and $n = 200$ denotes the number of wave crests that were, on average, included in a 30 minutes sequence. The terms $\alpha_w$, $\beta_w$ and $\gamma_w$ are the scale, slope and location parameters of the Weibull distribution, respectively, and were estimated through a least square method. The probability of occurrence for the characteristic largest event is 
\begin{equation}
\label{extr_gumbel}
F_E (H_{cr, max}) = 1-\frac{1}{n}.
\end{equation} 
Therefore, by incorporating equation (\ref{extreme_fun}) into (\ref{extr_gumbel}), the maximum crest can be expressed as
\begin{equation}
\label{extr_ch}
H_{cr, max} = \gamma_w + \alpha_w \left[ \text{ln}(n) \right]^{1/\beta_w}.
\end{equation}

\bibliography{apssamp}

\end{document}